\documentclass[10pt, a4paper, twocolumn]{article}
\usepackage[utf8]{inputenc}
\usepackage{microtype}
\usepackage{graphicx}
\usepackage{booktabs} 
\usepackage[prologue, table]{xcolor}
\PassOptionsToPackage{hyphens}{url}
\usepackage[hidelinks]{hyperref}
\usepackage{floatrow}
\usepackage{subcaption}
\usepackage{tikz}
\usepackage{enumitem}
\usepackage{authblk}
\usepackage{multicol}

\begin{document}
\title{Towards Understanding First-Party Cookie Tracking in the Field}
\author[1,3]{Nurullah Demir}
\author[1]{Daniel Theis}
\author[1,2]{Tobias Urban}
\author[1]{Norbert Pohlmann}
\affil[1]{Institute for Internet Security---if(is)}
\affil[2]{secunet Security Networks AG}
\affil[3]{KASTEL Security Research Labs, Karlsruhe Institute of Technology}
\date{
    \large
    \color{blue}
    \textbf{
        This is an extended version of our work that appeared  at Sicherheit'22}
}


\rowcolors{2}{gray!25}{white}

\microtypecontext{spacing=nonfrench}
\widowpenalty1000000
\clubpenalty1000000

\floatplacement{table}{tb}
\floatplacement{figure}{tb}

\newcommand{\etAl}{et\,al.~}
\newcommand{\eg}{e.g.,~}
\newcommand{\ie}{i.e.,~}

\newcommand{\empirical}[1]{#1} 

\newcommand*\rectangled[1]{%
      $\tikz[baseline=(R.base)]\node[draw,rectangle,inner sep=0.5pt, outer sep=0pt, fill=black, text=white](R) {#1};\!$
}

\maketitle


\begin{abstract}
    Third-party web tracking is a common, and broadly used technique on the Web. 
    Almost every step of users' is tracked, analyzed, and later used in different use cases (\eg online advertisement). 
    Different defense mechanisms have emerged to counter these practices (\eg the recent step of browser vendors to ban all third-party cookies). 
    However, all of these countermeasures only target third-party trackers, and ignore the first party because the narrative is that such monitoring is mostly used to improve the utilized service (\eg analytical services).
    
    In this paper, we present a large-scale measurement study that analyzes tracking performed by the first party but utilized by a third party to circumvent standard tracking preventing techniques (\ie the first party performs the tracking in the name of the third party).
    We visit the top 15,000 websites to analyze first-party cookies used to track users and a technique called ``DNS CNAME cloaking'', which can be used by a third party to place first-party cookies.
    Using this data, we show that \empirical{76\%} sites in our dataset effectively utilize such tracking techniques, and in a long-running analysis we show that the usage of such cookies increased by more than 50\% over 2021.
    Furthermore, we shed light on the ecosystem utilizing first-party trackers, and find that the established trackers already use such tracking, presumably to avoid tracking blocking.
    
\end{abstract}

\section{Introduction}
\label{sec:intro}
The business model of many modern (Web) applications relies on the revenue generated by ``renting'' space on their services to advertisement companies.
These ad-tech companies try to place advertisements on the sites that meet the users' interests motivating that they will interact with the ad, and ultimately buy the advertised product or service.
To place such targeted ads, ad-tech companies track users across the Web, by assigning a unique identifier to each of them, and try to understand their interests by building so-called \emph{behavioral profiles}~\cite{McDonald.2010}.
These profiles are built based on many datat sources (\eg the users' clickstream) the browsing or usage behavior of the user and might even include detailed information like a click history in order to get more information what a user is doing within a specific site.
The unique user identifiers are often stored in the third-party context (\eg in an HTTP cookie).
Some consider this large-scale tracking as privacy-invasive because it often happens without users' explicit consent or knowledge~\cite{truste2017}, nor is the tracking made transparent to the user.
The desire for more privacy and the need for more user data led to an arms race between anti-tracking tools, and novel techniques to track users.
One recent (technical) step in this race was the announcement of major browser vendors to ban third-party cookies within the next years~\cite{Google.Cookies.2020,Mozilla.Cookies.2020}.
While there is no immediate problem with third-party cookies, previous work showed that they are overwhelmingly used for advertisement purposes~\cite{Urban.WWW.2020}.
Hence, trackers need to find different ways to persist their identifiers on the users' devices.
One known way to do so is the computation of \emph{browser fingerprints}, which are distinct identifiers that are computed based on properties of the user's device or browser~\cite{Englehardt.OpenWPM.2016,Laperdrix.FP.2020}.
However, these fingerprints change over time and, therefore, one cannot simply rely on them for tracking purposes~\cite{Gomez.Fingerprint.2018}.

Another way to cache such identifiers is to store them in a first-party context, and send them to a third party if needed.
More specifically, a tracking script  embedded in the first-party context of a site could store the identifier (cookie) in the first-party context and send it to the tracker in a dedicated request.
Hence, deleting, or banning third-party cookies does not affect them.
In this paper, we perform a large-scale measurement study on the top 15,000 sites to analyze the presence of such first-party tracking techniques in the field and use the HTTP Archive~\cite{HttpArchive.2021} to analyze the development of such tracking techniques.
In our measurement, we find that roughly \empirical{90\%} of all sites include a first-party object that tracks users, and leaks the identifier to a third party.
Furthermore, our results show that \empirical{10.354 (69\%)} of the sites in our analysis corpus hide the presence of a tracker by redirection of request on DNS level (``DNS CNAME cloaking'').
By analyzing the players that use these techniques, we find that the ones who already play a dominant role in the tracking ecosystem (\eg\emph{Google} and \emph{Facebook}) dominantly utilize them.
Using the HTTPArchive~\cite{HttpArchive.2021}, we show that first-party tracking, in cooperation with a third party, has been a common phenomenon in the past with a growing popularity. 
Our results show that in 2021 the leakage of such cookies to a third party increased by nearly 50\%.

Previous work that analyzed the tracking ecosystem almost exclusively focused on third-party trackers from various perspectives (\eg\cite{Acar.Tracking.2014,Englehardt.2015,Englehardt.OpenWPM.2016,Gomez.Fingerprint.2018,Fouad.Missed.20}). 
Other works focused on analyzed tracking attempts performed in the first party context using CNAME cloaking~\cite{Dao.CNAME.2020, Dimova.CNAME.21, Ren.CNAME.21} (see Section~\ref{sec:background}).
CNAME cloaking is one way to pace for a third party to place an identifier in the first-party context of a site (see Section~\ref{sec:cname_tracking}).
In this work, we focus on first-party tracking via cookies and do not aim to analyze tracking enabled by CNAME cloaking, but we see it as one way to place such cookies.
One important insight from our work is that tracking is no longer a phenomenon only present in the third-party context but has arrived in the first-party context at large scale.
Consequently, many of the current anti-tracking tools might have to be revised to match this new tracking scope, that future studies might have to reconsider their measurement setups, and that banning third-party cookies might have less impact than it is currently expected~\cite{party-cookies-gone.2020}.

In summary, we measure the extend and effects of first-party tracking in the field, and make the following key contributions:
\begin{enumerate}
    \item We perform a large-scale web measurement of the top 15,000 sites on the Web, and analyze if they utilize first-party identifiers. Our results suggest that first-party tracking -- at the behest of a third party -- arrived at scale on the Web with a significant impact on users' privacy.   
    
    \item In our experiment, over \empirical{85\%} of the analyzed sites store potential tracking identifiers in a first-party cookie, and send them to a third party. \empirical{20\%} of all identified (first-party) cookies store an identifier that can be used to track users, and we found that \empirical{69\%} of the analyzed sites use a tracker that is cloaked on DNS level. Utilizing the \emph{HTTPArchive}, we show that this technique has already been used in 2019 and has ever grown since.

    \item Finally, we analyze the companies participating in the first-party tracking ecosystem. We find that trackers who dominate the third-party tracking market also dominate the first-party tracking market. However, we observed differences in companies that utilize different first-party tracking techniques.
\end{enumerate}

\section{Background}
\label{sec:background}
Web tracking usually describes the practice of website operators to integrate a tracker into their websites and capture the browsing behavior of their visitors. Trackers allow to collect data about the website visitors wherever the same tracker is integrated. For instance, a wide spread tracker collects a set of sites or a browsing profile that resembles the browsing behavior of website visitors.

Over the course of the last years tracking became an integral component of the web, therefore, so called third-party trackers can be found in the vast majority of websites ~\cite{Ler16, Urban.WWW.2020, Englehardt.OpenWPM.2016, Englehardt.2015}. The idea is that a third party provides a tracking mechanism to a broad variety of website operators which integrate the tracking code into their websites. 
The reasons to track users ranges from site analytic to cross-domain tracking with the means to provided targeted ads.

The ratio for this are desirable features like site analytics, personalization and targeted advertising, basically means to understand the website visitor respectively customer better.
The larger the set of websites a tracker covers the better is the service it can provide to its customers (the website operators) and in theory also to the visitor in the form of personalization. According to Roesner et al.~\cite{Roesner.2012} a tracker can serve multiple functions (\eg analytics and tracking) at once and also work within different scopes (cross-site or within-site).

Tracking can be realized in a multitude of ways for example by the use of web beacons, browser fingerprinting, or CNAME cloaking~\cite{Sim15, CNAME.Hackernews.2021}. 
Often tracking techniques need to store a unique identifier on a system (\eg HTTP cookies).
In this paper, we focus on  tracking methods via so called ``first party cookies'', which are usually created by a web server ~\cite{Cas12, Iet11, Onp20}. 
In essence, cookies are textual data that is stored  on the client by a server, and set via HTTP requests.
Cookies can either be temporary or persistently stored.
Temporary cookies are called session cookies or transient cookies which store information about the current session (\eg authentication data). 
Persistent cookies, on the other side, are stored until they expire -- which can span from a day to several months -- or are specifically deleted (by the user) and store data that shall persist throughout multiple sessions like settings.
Each HTTP responds includes the cookies set by the specific domain.
If such a cookie contains an identifier it allows the operator of the domain to track users~\cite{Cas12}.
When a cookie is set a returning user respectively visitor sends their cookie(s) to the server, which allow the operator to track the visitor~\cite{Cas12}.

Cookies can be categorized into two mayor contexts: first and third-party cookies. 
\emph{First-party cookies} are set by the server of the visited website and are usually used to enhance the user experience by storing data related to site interactions like settings (\eg preferred language) or shopping carts~\cite{Sim15}. 
In contrast, a cookie is called a \emph{third-party cookie} if the visited website embeds an object from another domain and this third party sets the cookie.
Such ads may use the cookies to enable targeted advertising and allow profile building for the tracker and the website operator. 
Such third party cookies are often used  to store user identifiers, especially by ad-tech companies, party cookies, to track users across the Web~\cite{Sit01, Tou12, Urban.WWW.2020}.
To oppose these threats browsers ~\cite{Web21, Moz21}, browser addons and other tools aim to block especially third-party cookies to decrease the privacy impact on users~\cite{Boh20, Google.Cookies.2020, Mozilla.Cookies.2020}.
However, these attempts only focus on the third-party context leaven a blank spot on the first-party side.
Therefore, the risks for the privacy on individuals definitely is imminent, yet a few years ago a development started from governments and the industry towards a more privacy friendly interaction between legal entities and natural persons. With this current development of private companies~\cite{Web21, Moz21} and public institutions~\cite{Eur16} regulating or even inhibiting the usage of cookies a new development started where website operators mis- or abuse first party cookies as third party cookies in order to still allow for tracking their visitors.

In the following subsections we introduce the tracking model analyzed in this paper, the means to store data in a browser context as well as canonical name cloaking.
Throughout the paper, we use the term \emph{site} if we talk about the registerable part of  a domain (\ie the eTLD+1) and the term \emph{page} to describe a specific HTML document on a site.
An example for a site could be \emph{foo.com} and a page on that site could be \emph{https://foo.com/news}.

\subsection{First-Party Tracking Model}
\label{sec:attack_model}
\begin{figure*}[tb]
    \centering
    \includegraphics[width=\textwidth]{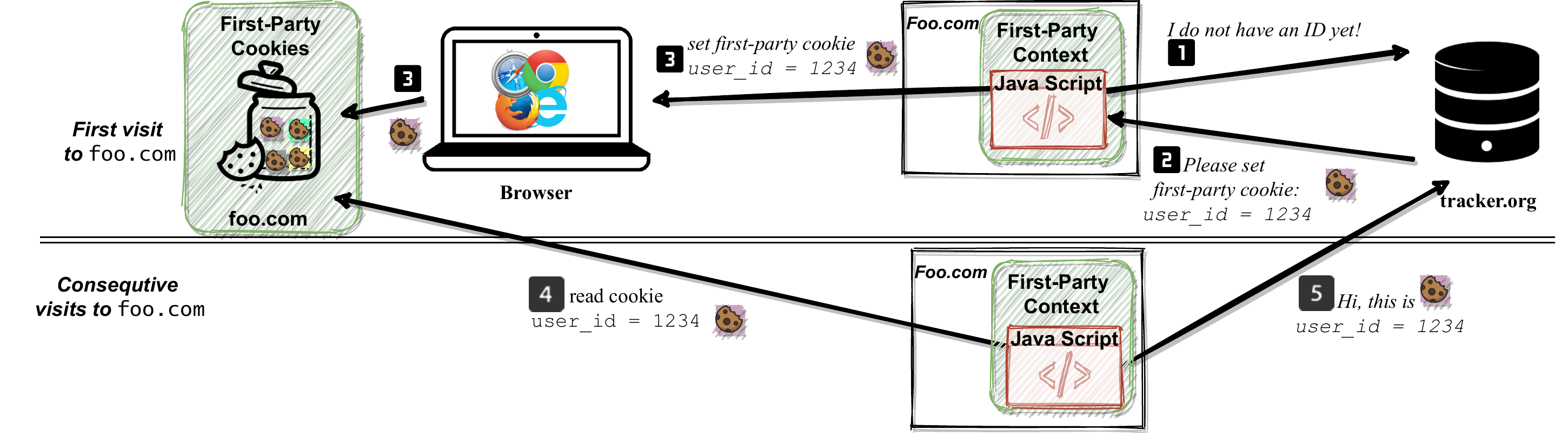}
    \caption{Companies active in the first-party tracking ecosystem.}
    \label{fig:attack_vector}
\end{figure*}
In contrast to traditional third-party tracking, in which the first party (website) does not necessarily know which trackers might be embedded into the website~\cite{Urban.WWW.2020}, in our tracking model, the first party needs to support the tracker actively.
The high-level idea is that the first party stores the tracking information in its first-party context and then sends it to a third party.
Figure~\ref{fig:attack_vector} displays the tracking (attack) model that we assume in this work.
Our model consist of three parties: (1) the tracker (\texttt{tracking.org}), (2) the visited website (\texttt{foo.com}), and (3) the users' browser.
More specifically, the website places a (static) interactive object (\eg a JavaScript code) in its first-party context that establishes a connection to a known third-party interface (\eg by issuing an HTTP GET request).
Once the user visits a website, the first-party object checks if an identifier for the user exists in the local first-party context (see Section~\ref{sec:first_party_storage}).
If it does not exists, a new identifier is set (\rectangled{1} in Figure~\ref{fig:attack_vector}).
This identifier can either be computed based on the device's properties (\ie a device fingerprint) or be generated by the third-party (\rectangled{2}).
The browser will store the identifier in the first-party context of the site since it was set by a first-party object (\rectangled{3}).
On consecutive visits the first-party tracking object can read the previously set tracking identifier (\rectangled{4}) and send it to the third party (\texttt{tracking.org}, \rectangled{5}).
Utilizing this mechanism, the tracker can bypass defense mechanisms that aim to limit or eliminate third-party cookies.  

Henceforth, We call this practice potential ``\emph{ first-party cookie tracking}'' because it works like traditional tracking only that the identifier is stored in the first-party context.
This method comes with the drawback that the identifier is not accessible cross-site per-se since it has to be set for each site and might be accessed by other third parties that get access to the storage.
Naturally, an identifier that is stored by the first-party cannot be used by a third party to track the user.
However, once the identifier is leaked to a third party, it can potentially be used by that third party to track the user across a site.
This could be the case for consent management services that assign a (first-party) identifier to a user, and, consequently, track them across the site.
In this case, tracking is done with an eligible purpose (consent management), but the third party \emph{could} also abuse it.

\subsubsection{Means to Store Data}
\label{sec:first_party_storage}
Modern web browsers offer various means to store  data in the first party context.
Two popular ways are the browser's \emph{HTTP cookie jar} and \emph{local storage}, which was introduced in HTML5.
Cookies are the legacy option to persistently store data based on a single string. This is done at least for each individual domain. Moreover, although limited to a size of 4kB, cookies are sent unencrypted and may therefore pose a risk for privacy and security. On the other hand \emph{local storage} is an alternative to persistently store data directly in the browser. This method of storing is only supported by modern browsers, it is not sent with every \emph{HTTP Request} and allows to store up to 5MB of data per domain and, moreover, has other than cookies no expiration limit.
In the remainder of the paper, we will use the term \emph{cookie jar} as a synonym for all previously described means to store data in the first-party context.

\subsubsection{Canonical Name Cloaking}
One popular way to link a third-party resource with a first-party element is to use so called \emph{CNAME cloaking}.
Within the DNS protocol, the \emph{canonical name record} (CNAME) is used to map a domain name to another.
This mapping is commonly used to host multiple services (\eg \emph{news.foo.com}, \emph{ftp.foo.com}, and \emph{mail.foo.com}) on the same IP.
To do so, one creates an alias (CNAME) for each service that all refer to the same \texttt{DNS A} record of \emph{foo.com}.
However, the CNAME can also point to another server.
For example, the CNAME of \emph{news.foo.com} could point to \emph{bar.com}.
On the Web, this means that the browser will load the content from \emph{bar.com}, and not \emph{foo.com}.
This technique is known as \emph{CNAME cloaking}.
CNAME cloaking is commonly used on the Web.
For example, content delivery networks might use a CNAME cloaking to run an application (\eg \emph{foo.com.cdn.org}) which is still accessible via its original address (\eg \emph{foo.com}), which registers a CNAME for the CDN.
However, one can also exploit this to avoid that users block specific content (\eg ads or tracking) as this is commonly done based on the loaded URL~\cite{Merzdovnik.2017}.
For example, if a user wants to avoid content loaded by \emph{tracker.com} and blocks all requests to the domain on the application level (\eg by using an ad-blocker), a web service could circumvent that by resolving all requests to \emph{t.foo.com} to the tracking domain -- on DNS level.

\section{Method}
\label{sec:method}
In this section, we provide an overview of our measurement framework (see Section~\ref{sec:measurement_framework}) and the steps we take to measure first-party tracking attempts in the wild (see Sections~\ref{sec:measure_cname} and~\ref{sec:measure_1st_cookie}).

\subsection{Website Corpus}
We chose to use the \emph{Tranco} top 15,000 list generated on 06/16/2020 (ID: PX8J)\footnote{Available at https://tranco-list.eu/list/PX8J.}, which is an aggregation of other popular top lists~\cite{LePochat.Tranco.2019}, in our analysis.
We visited each site on the list and collected 30 distinct first-party hyperlinks to pages on the same site, which we used in our measurement crawls. 
If possible, we repeated the process to recursively collect up to 30 subsites.
We chose to visit 30 pages instead of only analyzing the landing page of a site because recent work has shown that ``subsites'' show increased usage of privacy-invasive technologies and that visiting 30 sites is a good approximation to attest the privacy impact of a website~\cite{Urban.WWW.2020,Aqeel.Landing.2020}. 
Due to this need for a vertical measurement setup, the number of distinct websites we analyze is limited compared to other measurement studies (\eg\cite{Englehardt.2015}).

\subsection{Measurement Framework}
\label{sec:measurement_framework}
To measure the usage of first-party tracking and the leakage of such cookies, we utilize the popular \emph{OpenWPM} framework~\cite{Englehardt.OpenWPM.2016}, which uses the Firefox browser to visit websites.
We configured the framework to log all (1) HTTP requests/responses and (2) data stored in the local storage or cookie jar.  
Furthermore, we instrumented Firefox's DNS API~\cite{DNS.Mozilla.2020} to log the DNS resolution of all hostnames, which is necessary to identify CNAME cloaking (see Section~\ref{sec:measure_cname}).
For each visited page, we configured the framework to use the same user agent (\texttt{Mozilla/5.0 (X11; Ubuntu; Linux x86\_64; rv:77.0) Gecko/20100101 Firefox/77.0}) and desktop resolution (\texttt{1366x768}) to ensure that these attributes will not impact the computation of browsers fingerprints.
According to Urban~\etAl\cite{Urban.WWW.2020}, websites tend to use more cookies if the cookie jar and local storage of a browser instance is populated.
Therefore, we created a base profile that is loaded by \emph{OpenWPM} upon each page visit.
To populate this profile, we successively visited each landing page in our dataset once and used the resulting Firefox profile in our measurement.
The base profile is not updated after each visit (\ie we perform a \emph{stateless} crawl), which, on the one hand, implies that the order of visited websites has no impact on the results and, on the other hand, that trackers have to re-set their identifiers.
Hence, this setup allows us to distinguish between user identifiers and session IDs.

We conducted the measurement from a university network within the EU.
To perform the measurement, we profit of two machines which in total have 95\,GB of RAM, 56 CPUs (\emph{Intel Xeon CPU @ 2.10GHz}), and more than 5\,TB of hard disk space.
When visiting a website we used a 30 second timeout interval to compensate for \eg site crashes or slowly or not loading websites.
We did no retry a page visit in case it failed in the first attempt.
In our crawl, we used simple faked user interactions (\ie random scrolling and mouse jiggling) because previous work found that this increases the number of observed tracking attempts~\cite{Urban.WWW.2020}.

In addition from the already named browser configuration, we use four different browser configurations (``profiles'') for our measurement: (1) No particular configuration, (2) disabled 3rd party cookies, (3) active anti-tracking tool (\ie \emph{uBlock Origin}~\cite{uBlock.2020}; henceforth, ``\emph{uBlock}''), and (4) disabled 3rd party cookies, and an active anti-tracking tool (\ie (2) and (3) combined).
We chose to use the option to block third-party cookies to test if trackers use other techniques (\eg first-party tracking techniques) if they cannot store their identifiers as they usually do.
We used \emph{uBlock} because -- to the best of our knowledge -- it is the only tool at the time of writing this paper that claims to identify and block CNAME cloaking.
Hence, the tool provides protection against trackers that utilize this new technique, and we can assess, concerning the other profiles, how effective it is.
While the named defence mechanisms do not actively aim to protect against tracking performed by a third party they might block requests to a third party that contains the first-party identifier, which would effectively limit the tracking impact.
For each profile, we performed a separate measurement crawl.
Hence, we visited each page on our website corpus four times, once with each profile.

Like most large-scale web measurement studies, our work comes with the limitation that our crawler is not logged in to any website, nor do we perform any meaningful interaction with the websites.
Hence, we will miss some pages and features of sites that are only triggered after a user interaction (\eg payment processes) or after successful login  (\eg loading of personal data).
Regarding tracking technologies, this means that some sites that wait until the user opts-in, might not set any cookies.
Since the literature suggests that these opt-in choices (``cookie banners'') are often not honored~\cite{Sanchez.OptOut.2019,Utz.Consent.2019}, our results can only be seen as a lower bound to the extend of the problem.
Furthermore, our automated crawler might be detected by a page, which might then show different content than for real user visits.

In our analysis, we use the \emph{WhoTracks.Me} database to map domains (eTLD+1) to the organizations running them~\cite{WhoTracksMe2018}. 
This clustering allows us to discuss the impact of a company rather than the impact of a domain because one company can run multiple tracking domains.
Previous works have shown that some types of websites tend to use more (third-party) user tracking techniques than others (\eg ``News'' websites)~\cite{sorensen2019before}.
Therefore, we analyze if similar observations can be made for first-party tracking techniques as well.
We use the \emph{McAfee SmartFilter Internet Database} service to identify the primary purpose (``category'') of a domain~\cite{mcaffee.2019}.
This mapping allows us to understand which categories of sites utilize the technique of interest and gives us a brief impression of who mainly uses it.
The \emph{Cookiepedia}~\cite{cookiepedia.2019} is a crowd sourced index for cookies, their purposes, and ``owning'' companies.
In this work, we use the service to map identified cookies to the respective companies who ``own'' and use them.

\subsection{Measuring CNAME Cloaking}
\label{sec:measure_cname}
CNAME cloaking is increasingly used to track users~\cite{Dao.CNAME.2020,Dimova.CNAME.21}. 
We analyze this practice's presence by inspecting the DNS resolutions on the client by instrumenting the DNS resolution of the Firefox browser.
More specifically, we log the URL handled by the browser and the responding DNS resolution observed by the browser, which we instrumented.
For example, if we observe a request to \emph{tracking.foo.com/search=foo} whose domain is then cloaked to \emph{cloaked-party.com}, we use the URL \emph{cloaked-party.com/search=foo} in our analysis but treat it a first-party request.
In contrast to previous work in the field of CNAME cloaking field, this allows us to analyze the real DNS resolution of each request.
Previous work often performed the DNS resolutions isolated from the web measurement on different machines and at different times~\cite{Dimova.CNAME.21, Dao.CNAME.2020}.
This comes with the non-negligible assumption that DNS resolutions will not change over time and/or based on the location of users, which could change the outcome of a study.
However, we want to highlight that our contribution is not the measurement of tracking that is enabled by third parties but that we are interested in tracking identifiers that are stored in a third-party context.
One way to store such cookies is via CNAME cloaking.

In order to compare our results to previous work, we match the resolved URL against two standard tracking filter lists, as of 04/16/2021 (\ie the \emph{EasyList} and \emph{EasyPrivacy} lists ~\cite{EasyList.2019}).
Such lists always come with the downside that they might not contain all tracking URLs~\cite{Fouad.Missed.20,Merzdovnik.2017} and, therefore, our results can be seen as a lower bound.
However, since we used lists generated after the crawl, we think they should contain most trackers.
As already mentioned, CNAME cloaking is not a problem per se, but it could be used to hide the real recipient of a request.
For example, a service provider who wants to host a domain on a content delivery network but wants to keep his domain (brand) name often uses CNAME cloaking to do so~\cite{Dao.CNAME.2020}.
To preclude cloaking instances where that redirect to very similar domains, we performed a simple similarity comparison of the original hostname and the cloaked one.

As previously described, we say that the CNAME of a request is cloaked if it is resolved to another domain (\eg\emph{foo.com} is redirected to \emph{bar.com} on DNS level).
We analyze this practice's presence by inspecting the DNS resolutions on the client by instrumenting the DNS resolution of the Firefox browser.
More specifically, we log the URL handled by the browser and the responding DNS resolution observed by the browser, which we instrumented.
Hence, we have access and can analyze the domain part present on browser level and the one to which the request is sent, which is usually not visible in the application.
We then substitute the domain part of each URL with the resolved domain name and leave everything else in place.
For example, if we observe a request to \emph{tracking.foo.com/search=foo} whose domain is then cloaked to \emph{cloaked-party.com}, we use the URL \emph{cloaked-party.com/search=foo} in our analysis.
To preclude cloaking instances that redirect to very similar domains, we performed a simple similarity comparison of the original hostname and the cloaked one.
If both are more than 70\% similar, we assume that it is the same domain and that no cloaking happened.
For example, if we observe that a request to \emph{www.google-analytics.com} is cloaked to \emph{www-google-analytics.l.google.com} we ignore it.
As already mentioned, CNAME cloaking is not a problem per se, but it could be used to hide the real recipient of a request.
For example, a service provider who wants to host a domain on a content delivery network but wants to keep his domain (brand) name often uses CNAME cloaking to do so~\cite{Dao.CNAME.2020}.
However, if a domain is cloaked that performs actions not desired by the user (\eg tracking or providing malicious content), it will most likely happen unnoticed by her.
To identify whether a cloaked CNAME is used for tracking purposes, we match the resolved URL against two standard tracking filter lists, as off 09/10/2020 (\ie the \emph{EasyList} and \emph{EasyPrivacy} lists ~\cite{EasyList.2019}).
In our analysis, we match the substituted URLs against the two lists ~\cite{EasyList.2019}.
Such lists always come with the downside that they might not contain all tracking URLs~\cite{Fouad.Missed.20,Merzdovnik.2017} and, therefore, our results can be seen as a lower bound.
However, since we used lists generated after the crawl, we think they should contain most trackers.

\subsection{First-Party Cookie Tracking}
\label{sec:measure_1st_cookie}
Since browser vendors limit the use of third-party cookies~\cite{Google.Cookies.2020,Mozilla.Cookies.2020}, trackers need to find different means to store them.
One way to store them is to use first-party cookies because these are currently not restricted.
In our work, we identify such ``1st-3rd-party cookies'' by analyzing the cookie jar and local storage and test if any of these cookies could potentially hold an identifier and if it is sent to a third party.
One challenge in this approach is that we have to distinguish between session cookies, which are usually not stored in third-party cookies, and user identifiers.
If the same identifier reoccurs on different visits to pages of the same site, we assume that it is a user identifier and not a session ID.
This assumption is valid because we perform a stateless crawl, and session IDs will reset between two visits since the local storage and cookie jar are reset. 
We assess a cookie to be a first-party cookie if it was set by a request or script that was loaded in a first-party context. 
Hence, if we visit \emph{foo.com} and the site sets a cookie, we count it as a first-party cookie. 
However, let us assume \emph{bar.com} embeds an object from \emph{foo.com} that sets a cookie; we do not count this cookie in our first-party analyses -- because it was set in a third-party environment.

In theory, cookies are simple \texttt{name=value} pairs, yet Gonzalez \etAl show that a single cookie often contains multiple values in proprietary formats (\eg \texttt{key=[v1= foo;v2=bar]})~\cite{gonzalez2017cookie}.
In our analysis, we try to URL decode all cookie values and split them at common delimiters (\eg \texttt{\_}, \texttt{=}, \texttt{:}, or \texttt{;}) and treat each token as a single value. 
Since this process might miss identifiers or inaccurately split them, our results can only be seen as a lower bound.
We apply the following definition for IDs to each of these values to test if they could serve as an identifier.
To identify tracking cookies, we use the definition of Koop \etAl\cite{Koop.2020}, which is inspired by the works of Englehardt and Acar \etAl\cite{Englehardt.2015,Englehardt.OpenWPM.2016,Acar.Tracking.2014} and commonly used to identify tracking cookies~\cite{Urban.Cookie.2020, Urban.WWW.2020}.
Following the definition, a cookie (value) can only be used to track users if and only if:
\begin{enumerate}
    \item it is not a session cookie and has a lifetime of more than 90 days;
    \item has a length of at least eight bytes (to hold enough entropy);
    \item is unique in each measurement run, and the length of each value only differs by up to 25\%; and 
    \item The values of the cookie are similar according to the Ratcliff/Obershelp~\cite{ratcliff.patter.1988} string comparison algorithm ($\le$60\%).
\end{enumerate}
We resort to theses definitions to allow basic comparison of our and previous work.

It is essential to mention that first-party cookies might hold an identifier to implement analytical or similar services, which are essentially a form of local user tracking. 
However, if these services send the (local) identifier to a third party, which offers the service, this third party could track the user.
For example, if we find the first-party cookie \texttt{OptanonConsent=ABC-123}, which is probably used by \emph{OneTrust}, and observe a third-party request that holds this ID (\eg\emph{tracking.foo.com?\textbf{consentId=ABC-123}}), we assume that \emph{tracking.foo.com} could potentially track the user with ID \emph{ABC-123}.
Hence, to find potential tracking that utilized first-party cookies, one has to find instances where they leak it to any third party.
In our analysis, we identify such cookie leakage by inspecting the HTTP GET and POST parameters of all third-party requests.
We compare them with all cookie values, that could be used for tracking (see the definition above). 
If a first-party cookie is leaked in such a way, we assume that it can be potentially used to track users.

\subsubsection{Cookie Tracking Over Time}
\label{sec:tracking_over_time}
The introduced methods potentially used to perform first-party tracking do not rely on any new or recently introduced techniques.
Hence, it is reasonable to assume that if such practices are used that they were used in the past.
To understand whether such practices were used in the past we utilize the \emph{HTTPArchive}~\cite{HttpArchive.2021}.
HttpArchive is an initiative to measure and analyze how the modern Web is built.
To achieve this, HTTPArchive crawls the landing pages of the most popular origins, based on the \emph{Chrome User Experience Report} (CrUX)~\cite{CrUX.2020}, on a monthly basis since 01/2019.
Thus, we can analyze the first-party tracking capabilities of all websites in the archive over time.
An important limitation of this approach is that it excludes all cookies set via JavaScript (\ie we only see cookies set via the HTTP protocol).

Due to the size of the archive, which includes millions of websites in each measurement, we decided to perform the analysis on a quarterly basis for 2020.
Hence, we use the following measurement points in our study: 01/20, 04/20, 07/20, 10/20, and 01/21.
In the analysis, we extract all cookie values from the raw data provided in the archive.
Using the raw data, we can access all necessary data that we need in our analysis to identify tracking cookies (\eg lifetime), which are not naively available in the rehashed data in the archive.
Similar to our approach in our active measurement, we first identify first-party cookies that could be used for tracking purposes (see Section~\ref{sec:measure_1st_cookie}) and then test if such cookies are leaked to a third party in an HTTP GET or POST request.
Using this approach, we can identify websites, third parties, and leaked cookies.

One difference between third and first party cookies that store an potential user identifier is that they might be used for analytical services, which is form of local tracing.
Analytical services track users as well but it is--- in theory---limited to the site on which the analytical service is run because the website provider wants to learn more about the way how users use the site.
However, if the used third-party analytical library is designed to send the user identifiers to an external server, which performs the actual evaluation of the user's actions, this third party can potentially track each user on the site. 
Hence, to find potential tracking that utilized first-party cookies one has to find instances where they leak to any third party.
In our analysis, we identify such cookie leakage by inspecting the HTTP GET and POST parameters of all request if they contain a cookie value from the local cookie store.
If a first-party cookie is leaked in such way, we assume that it can be potentially utilize by to track users.
For example, if we find the cookie \texttt{\_ga=GA1.2.\textbf{ABC.123}}, which is probably used by \emph{Google Analytics}, in the local cookie jar for \emph{bar.com} and observe the following request: \emph{https://www.google-analytics.com/r/collect?\textbf{\_gid=ABC.123}} we assume that \emph{google-analytics.com} could potentially track the user with  ID \emph{ABCD.1234} on \emph{bar.com}.

\section{Results}
\label{sec:results}
In this section, we present the results of our work.
More specifically, we analyze ``native'' (Section~\ref{sec:cookie_tracking}) and CNAME-based first-party cookie tracking tracking (Section~\ref{sec:cname_tracking})
Finally we provide an overview of the first-party tracking ecosystem (Section~\ref{sec:ecosystem}).

\subsection{General Overview}
\label{sec:overview}
First, we want to provide an overview of the measured dataset. 
We conducted all four measurement runs in three consecutive weeks (each measurement took approx. five days).
We conducted the first measurement (\#1---no particular configuration) on 07/20/2020, and the last measurement (\#4) ended on 08/12/2020.
Across all measurement, we visited \empirical{272,659} distinct pages on \empirical{11,471} sites.
In total, we visited \empirical{974,794} URLs across our four measurements.
The difference in visited sites can be explained in the fact that the top list contains sites that are not meant to be directly visited in a browser (\eg link shorteners, CDNs, and APIs) or sites that could not be reached within the timeout of 120 seconds.
In total, \empirical{1,290,509} cookies were set during these visits (\empirical{1,045,440} first-party cookies and \empirical{245,069} third-party cookies).
Across all measurements, we observed \empirical{37,172,846} instances of CNAME cloaking of which \empirical{783,917 (2\%)} ultimately resulted in tracking attempts, based on the classification of the resolved URL.

A summary of the measurement runs is given in Table~\ref{tab:overall_results}.
The drop in analyzed sites and, consequently, pages can be attested to sites no longer being available (\eg HTTP 404 errors).
According to the numbers, it seems that blocking third-party cookies has no effect on the usage of first-party cookies, but using an anti-tracking tool, \emph{uBlock} in our case, leads to a decrease in the usage of both types of cookies.
Furthermore, the tool seemingly helps to block CNAME cloaking attempts.
\begin{table*}
    \caption{Overview of all measurements. \small{$\ast$ Firefox deletes these cookies after creation.}}
    \label{tab:overall_results}
    \centering
        \begin{tabular}{cllrrrrrr}
        \toprule
        \rowcolor{gray!50}
        \# & Configuration & Date & \# Sites & \# Pages & 1\textsuperscript{st}-party c. & 3\textsuperscript{rd}-party c. & CNAME Cl.   \\
        \midrule
        
        1 & Plain browser                               &  07/20 & 11,471 & 272,659 & 408,509 & 155,193 & 14,493,533 \\
        
        2 & No 3\textsuperscript{rd}-party cookies &  07/26 & 10,368 & 246,493 & 355,208 &     ---\textsuperscript{$\ast$} & 11,017,643 \\
        
        3 & \emph{uBlock Origin} active                 &  08/01 & 10,203 & 230,327 & 139,115 &  89,876 &    5,829,786 \\
        
        4 & \#2 and \#3 combined                        &  08/06 & 10,153 & 225,315 & 142,608 &     ---\textsuperscript{$\ast$} &   5,831,884 \\
        
        \bottomrule
        \end{tabular}
\end{table*}

\subsection{First-Party Tracking Cookies}
\label{sec:cookie_tracking}
Currently, there is a trend that major browser vendors aim to abolish third-party cookies~\cite{Mozilla.Cookies.2020,Google.Cookies.2020}.
Hence, third parties need to find other ways to store user-related data (\eg user IDs).
One way to do so is to use first-party cookies (see Section~\ref{sec:attack_model}).
In our measurement, we identified, on average, \empirical{260,931} first-party cookies in each measurement.

\begin{figure*}
    \centering
    \includegraphics[width=\textwidth]{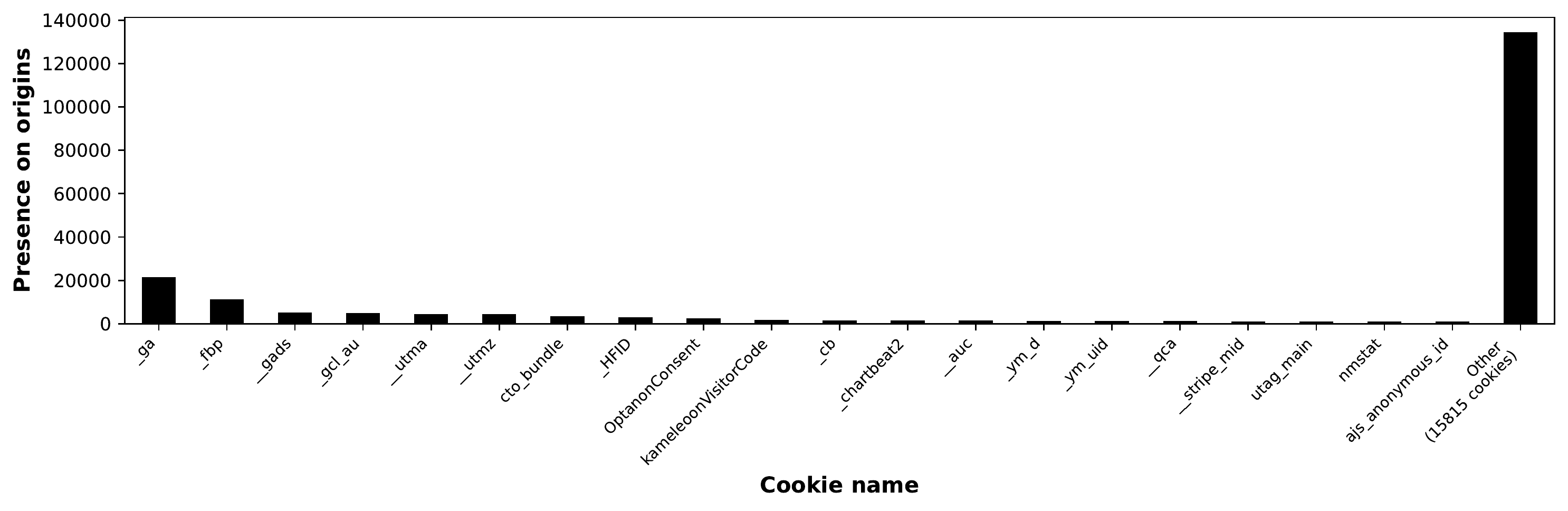}
    \caption{Number of origins that set a first-party cookie with the given name.}
    \label{fig:tracking_cookie_names}
\end{figure*}

Using our approach to classify tracking cookies, we found that, on average, each site uses \empirical{6} potential first-party tracking cookies (max: \empirical{598}; min: \empirical{0}) and over all profiles \empirical{19\%} of all first-party cookies are used to track users.
Overall, we found that \empirical{210.721 (20\%)} of all first-party cookies could be used to track users. 
In total, we found such tracking cookies on \empirical{10,700} (\empirical{93\%}) of the analyzed sites.
Hence, a vast majority in our dataset of websites offer third parties their first-party storage to place identifiers.
We used the name of a cookie as an indicator to distinguish which third-party utilizes the cookie.
The distribution of the names of these first-party cookies is given in Figure~\ref{fig:tracking_cookie_names}.
To increase the figure's readability, we only list the top \empirical{20} cookies, in terms of appearance.
We observed a long tail distribution, and to increase the readability of the figure, we only name the top 20 names, which cover over \empirical{45\%} of all tracking cookies.
The figure shows some prominent non-trivial cookie names used by several websites, which indicates that they all originate from commonly used services or libraries.
According to the classification of \emph{Cookiepedia}~\cite{cookiepedia.2019}, the top cookies, of the aforementioned classified cookies, are used by \emph{Google} (\texttt{\_ga} \empirical{28\%}), \emph{Facebook} (\texttt{\_fbp}, \empirical{14\%}), and again \emph{Google} (\texttt{\_\_gads}, \empirical{7\%} and \texttt{\_gcl\_au} \empirical{6\%}).
The results show that the large, known tracking companies are also utilizing first-party tracking tools, in addition to third-party tools.
Section~\ref{sec:ecosystem} provides a more detailed analysis of the first-party tracking ecosystem.
Compared to the results of previous work, that found that third-party cookies are almost exclusively used to track users~\cite{Urban.WWW.2020}, our results show that first-party cookies also serve other purposes (\eg preference management: \texttt{preferred\_locale=US}).

\begin{figure*}
    \centering
    \begin{subfigure}{0.45\textwidth}
        \includegraphics[width=\columnwidth]{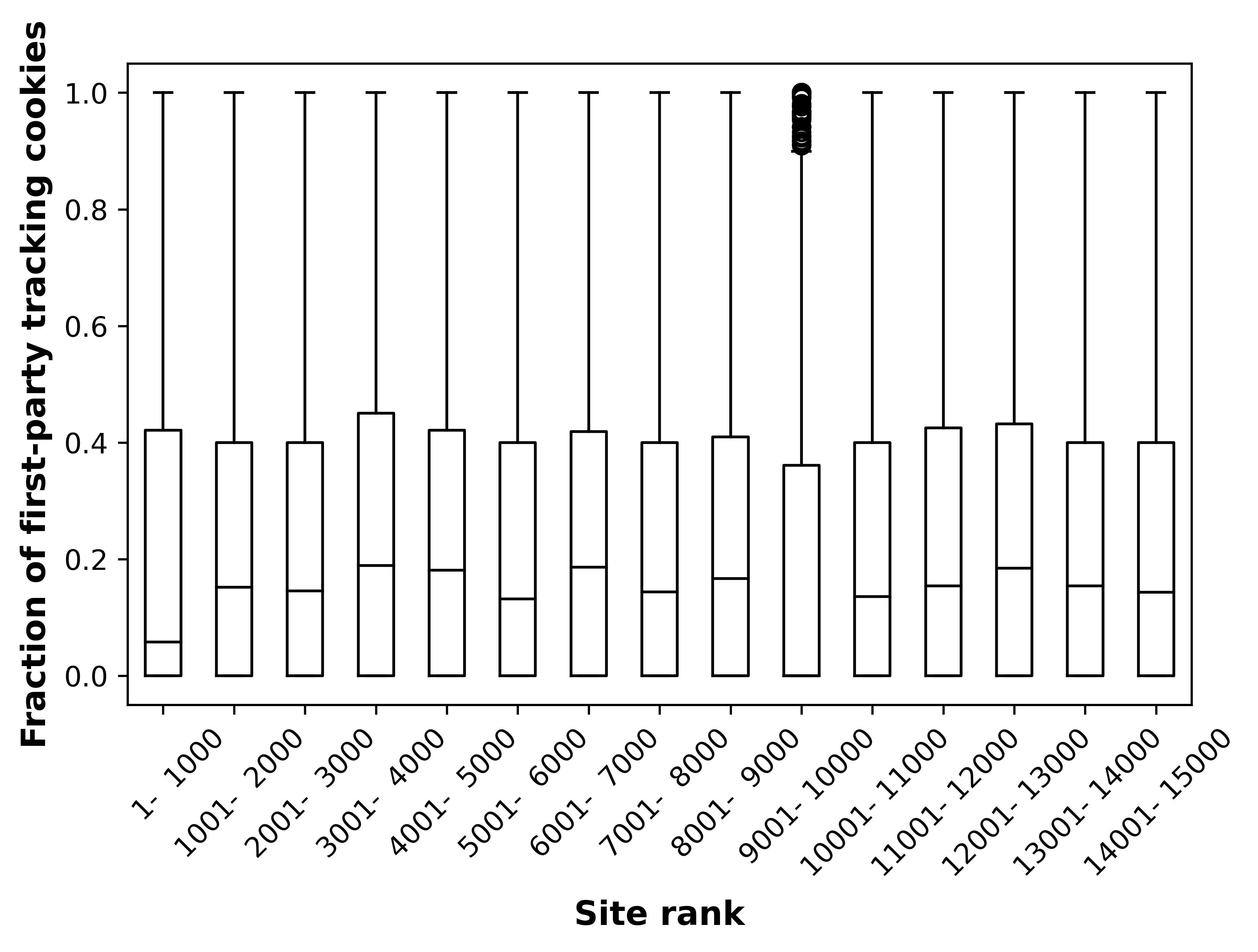}
        \caption{Profile \#1 and \#2.}
        \label{fig:frac_t1}
    \end{subfigure}
    \qquad
    \begin{subfigure}{0.45\textwidth}
        \includegraphics[width=\columnwidth]{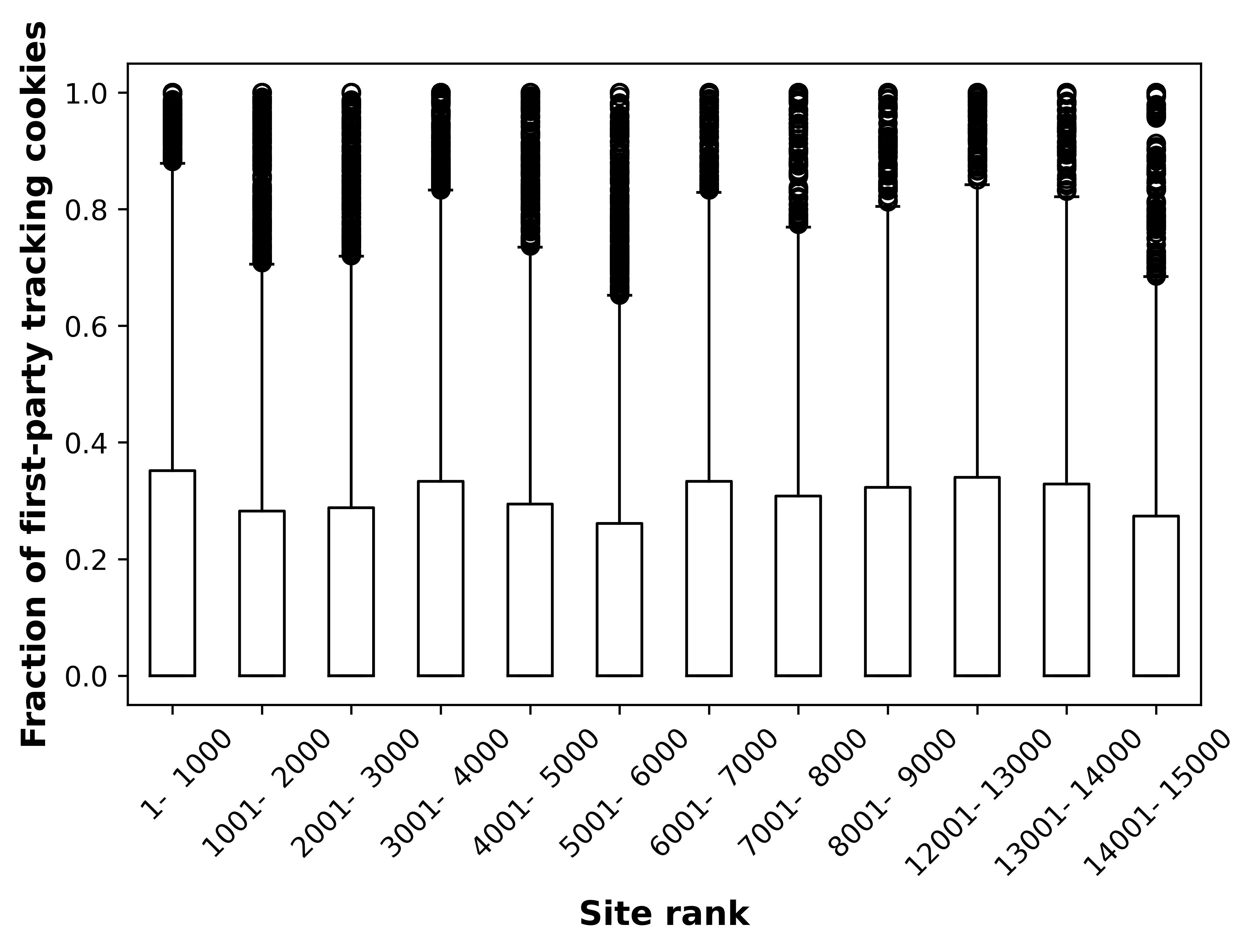}
        \caption{Profile \#3 and \#4 (\emph{uBlock active}).}
        \label{fig:frac_t2}
    \end{subfigure}
    \caption{Fraction of first-party cookies used to track users on the measured websites, by rank. Once if \emph{uBlock} is active (\#3 and \#4) and if not (\#1 and \#2).}
    \label{fig:frac_trackers}
\end{figure*}

In absolute numbers, popular websites (\ie sites with a low rank on the \emph{Tranco} list) tend to use more first-party cookies and also utilize more of them to track users.
In our experiment, popular websites (rank $\le 3,000$) use on average \empirical{45} cookies (standard deviation (SD): \empirical{65}) of which \empirical{21\%} (SD: \empirical{21}) are used to track users while less popular websites (rank $> 3,000$) use, on average, \empirical{34} (SD: \empirical{44}) cookies and \empirical{20\%} (SD: \empirical{61}) of them were classified as trackers. 
We chose the top 3,000 sites because less popular websites show a different behavior, number wise (see the standard deviations).
The $X^2$ test ($\alpha = 0.5$) also found a strong correlation between the absolute number of present trackers and the rank of the website ($p$-value $<0.001$).
Hence, such cookies are more likely to appear on popular sites.
However, if we look at the relative number of first-party tracking cookies in relation to other cookies, we find that popular websites use fewer trackers than other sites.
Figure~\ref{fig:frac_t1} shows the mean distribution of the relative share of first-party tracking cookies on the measured websites by their rank, which we observed in profile \#1 and \#2 (\emph{uBlock} deactivated).
One can see that the most popular websites utilize less first-party trackers, in relative numbers (median: \empirical{5\%}; SD: \empirical{31}), while other sites use a similar amount of such trackers (\empirical{14\%}; SD: \empirical{39}).
One takeaway from this finding is that first-party tracking via cookies is already a popular and well-established tool that is used in the field.
In comparison, Figure~\ref{fig:frac_t2} -- which summarizes these numbers from profile \#3 and \#4 -- shows that this distribution changes if \emph{uBlock} is activated.
We provide an in-depth analysis of the effect of \emph{uBlock} on the first-party tracking in the following.
To the best of our knowledge, most privacy-preserving tools do solely focus on analyzing third-party content to protect users, leaving them exposed to first-party tracking.

\subsubsection{Defense Mechanisms}
\label{sec:defense}
One simple defense mechanism to avoid third-party cookies all together is to tell the browser never to store them.
While third-party cookies are seemingly unrelated to first-party cookie sit is still interesting to analyses whether websites switch their tracking techniques if the preferred one is actively blocked (\ie do they store identifier in the first-party context if the third-party context is disabled).
In our measurement, this applies to profiles two and four (\ie we used Firefox's option to ``never'' store third-party cookies).
It is worth noting that the browser accepts and sets a third-party cookie but deletes it right away.
Hence, the script or request that sets the cookie can do this `normally' but cannot access the cookie afterward.
Overall, we observed a change of roughly \empirical{13\%} in the use of first-party cookies if third-party cookies are disabled.
This decrease can probably be attributed to the slight reduction of visited websites and the usual noise of large-scale Web measurements. 
Hence, blocking third-party cookies has a negligible effect on the use of first-party tracking cookies, which means that trackers do not use this way to evade the deletion of third-party cookies.
However, using \emph{uBlock} (Profile \#3 and \#4) shows that the usage of first-party cookies decreases by \empirical{65\%}. 

Overall, we found a decrease of active cookies that hold an identifier by \empirical{28\%} for profile \#2, \empirical{57\%} for profile \#3, and \empirical{51\%} for profile \#4. 
Hence, it seems that blocking third-party cookies is less effective than blocking entire requests to third-party if one wants to avoid \emph{first-party} tracking.
One explanation for this could be that the first-party tracking script initially loads content from a third party, using a \texttt{XHR} request, which includes the actual tracking code or essential parts of it.
This request might be blocked by an anti-tracking tool.

\subsubsection{Summary}
This section provides an overview of the extent of the utilization of potential first-party tracking cookies. 
We have shown that a variety of websites use them at scale.
Furthermore, the analyzed defense mechanisms seem to have a different impact on utilization. 
For example, blocking third-party cookies has almost no effect, while traditional tracking blocking has one.
However, while many first-party cookies hold values that can be used to track users, it is interesting to analyze how and if they are sent to a third-party, which could ultimately mean that they are abused as third-party trackers.

\subsection{First-Party Cookie Leakage}
As we have shown \empirical{20\%} of the first party cookies could be used to track different users across a site.
In this section, we evaluate to what extend these cookies are leaked to third parties.
A straight forward way to sent such cookies is to include them in HTTP GET or POST parameters.
As described in Section~\ref{sec:measure_1st_cookie}, we split the cookies values present in the local cookie jar at common delimiters and then look for the presence of the individual tokens in the observed requests.
Overall, we found that \empirical{68\% (143,216)} of the cookies that we identified as first-party trackers are sent to a single third party.
A common practice in the tracking ecosystem is the so-called \emph{cookie-syncing}, which means that two or more third parties exchange an user identifier.
In our experiment, we find that first-party tracking cookies are shared, on average, with  \empirical{1.3} third parties, across all profiles. 
\empirical{1,182 (0.8\%)} are shared with more than one third party.
Hence, real cross-domain tracking and ``cookie-snycing'' has not arrived in the first-party context at scale.
We identified \empirical{2,253} distinct domains that received such cookies, in \empirical{5,931,727} requests.
Our results show that first-party tracking cookies are often leaked to a third party and, therefore, they could be used to track users.
A more detailed analysis of the companies receiving cookies this way can be found in Section~\ref{sec:ecosystem}.

\subsubsection{Defense Mechanisms}
In profile \#3 and \#4, we enabled the \emph{uBlock} extension to test its effectiveness in protecting users.
The tool does not primarily aim to protect users' from cookie value leakage.
However, since it blocks some URLs that are knowingly used for tracking purposes it might also prevent first-party cookie leakage.
In our analysis, we observed \empirical{43\%} fewer requests that contained a cookie value.
This decrease consequently lead to a decrease of \empirical{98,819 (31\%)} tracking cookies leaked and \empirical{456 (20\%)} fewer third parties getting access to the cookie.
Similar to our findings on first-party tracking cookies, the leakage of them is also positively impacted by the used extension.
This is probably also explained by the fact that the tool blocks requests to different third parties

\subsubsection{First-Party Cookie Tracking Over Time}
Our active measurement shows that first-party cookie tracking and leakage are severe and joint problems on the Web.
In the following, we want to analyze the development of first-party cookie-based tracking usage.
To achieve this, we use data from \empirical{2019, 2020, and 2021} that we extracted from the \emph{HTTPArchive} (see Section~\ref{sec:tracking_over_time}).
The total size of the data set is \empirical{1.2}\,TB.

Across all \empirical{11} measurement points (four in 2019 and 2020 each and three in 2021), we analyzed \empirical{15.805.385} websites, which issued \empirical{5,592,914,767} requests.
Our dataset contains \empirical{119.676.680} first-party cookies.
Of those cookies \empirical{9,489,765} (\empirical{8\%}) could potentially be used to track users, according to our definition.
In total, \empirical{42,068} (\empirical{0.3\%}) of the websites actively sent a cookie to a third party, and \empirical{321.644} (\empirical{0.3\%} of the first-party cookies, and \empirical{3.4\%} of the tracking cookies)  were leaked in such away.

In the following, we only analyze sites that appeared in all measurement points.
The final dataset contains \empirical{1.777.206} websites, which issued \empirical{1,882,374,481} requests.
Those sites set \empirical{41,925,978} first-party cookies.
Of those cookies \empirical{2,668,340} (\empirical{6.4\%}) could potentially be used to track users, according to our definition.
In total, \empirical{13.116} (\empirical{0.7\%}) of the websites actively sent a cookie to a third party, and \empirical{63,235} \empirical{0.2\%} of the first-party cookies, and \empirical{2.4\%} of tracking cookies were leaked in such away.

We reason that this relatively high discrepancy to our active measurement is because of the large amount of less popular sites in the \emph{HTTPArchive} and our finding that popular sites tend to use first-party tracking than less popular ones.
When looking only at popular sites in the \emph{HTTPArchive}, we result that are comparable to our findings.
Over the analyzed period of almost \empirical{three} years, we see an increase in the relative number of leaked cookies of \empirical{99\%}  (01/19 vs. 07/21).
We compare relative numbers because not every measurement point in the HTTPArchive contains the same sites.
More specifically, some websites only occur in some crawls analyze more sites than others.
Hence, we cannot reasonably compare absolute figures. 
Furthermore, the number of third parties that receive such cookies increased by \empirical{36\%} and the number of websites that utilize them increased by \empirical{117\%}, in the same period.
These numbers indicate that, overall, first-party tracking becomes more and more popular on the Web.

\begin{figure}
    \centering
    \includegraphics[width=\columnwidth]{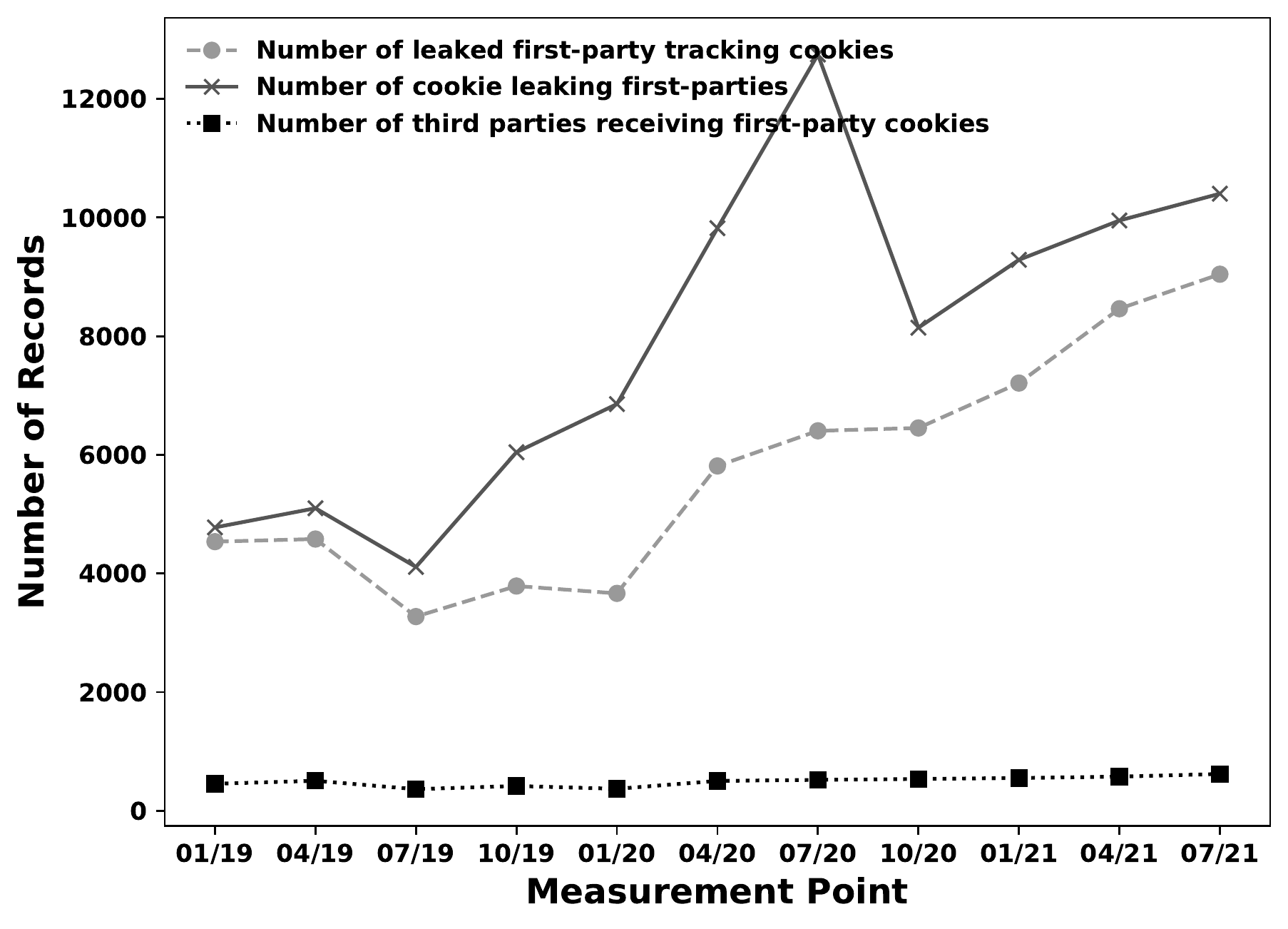}
    \caption{Number of distinct sites and third parties that engage in  first-party tracking and number of leaked first-party tracking cookies, over time.}
    \label{fig:first_party_tracking_over_time}
\end{figure}

Figure~\ref{fig:first_party_tracking_over_time} provides an overview of the temporal development of first-party cookie based tracking on the Web.
As one can see, the number of leaking parties and leaked cookies are directly related to each other.
This relation shows that a number of websites seem to use a more or less fixed set of services (\eg a consent management system) that use the first-party context to track users. 

Overall, the data point at 07/21 contains \empirical{118}\% more first parties that use the analyzed tracking cookies than the first measurement point (07/19).
The steep decline of such parties  between 07/20 and 10/20 is probably related to artifacts in the \emph{HTTPArchive} dataset.
The number of third-party trackers that receive such cookies remains relatively stable, with a slow increase from \empirical{456 (01/19)} to \empirical{621 (07/21)}, which accounts for an increase of \empirical{36\%}.
This development indicates that there is a steadily growing set of players in the ecosystem that utilize this technique (see Section~\ref{sec:ecosystem}), but more and more websites adopt these services of these players.
Appendix~\ref{app:over_time} shows an analysis of this development for all sites in the \emph{HTTPArchive} dataset and not only the ones that are present in all measurement points.

\subsubsection{Summary}
This section shows that the first-party storage of a site is often utilized by a third party.
This finding is an indication that third-party tracking is not only limited to the third-party context but also uses the first-party context.
However, traditional tracking protection mechanisms help to decrease the amount of data leaked to third parties.

\subsection{First-Party Cookie Tracking via CNAME Cloaking}
\label{sec:cname_tracking}
Previously, we discussed straight forward attempts of first-party cookie tracking, which presumably happens in coordination with a third party.
Another option for a third party to set and access cookies in a first-party context is that the first party redirects requests on DNS level to another domain (``CNAME cloaking'' -- see Section~\ref{sec:measure_cname}). 
Previous work has shown the extend of CNAME based tracking in the wild~\cite{Dao.CNAME.2020,Ren.CNAME.21,Dimova.CNAME.21}.
In this work, we show that this technique is also present in the first-party cookie tracking ecosystems, and we replicate high level  results of previous work to highlight the presence of this novel tracking technique.

Utilizing CNAME cloaking, HTML objects (\eg JavaScript) operate in the first-party context (\eg \emph{t.foo.com}) on browser level but are actually loaded from another domain (\eg \emph{tracking.org}) on DNS level.
Across all four measurements in our dataset, we found over \empirical{37M} requests for which the DNS name was cloaked (\empirical{45\%} of all requests).
Of those requests \empirical{2.1\%} (\empirical{783,917}) are sent to an URL that is marked to be used for tracking purposes, by the lists mentioned in the Section~\ref{sec:measure_cname} \emph{after} the cloaking of the CNAME.

Our results show that CNAME cloaking is commonly used but also abused to disguise trackers, which is in line with previous work~\cite{Dao.CNAME.2020,Ren.CNAME.21,Dimova.CNAME.21}.
To get a better understanding of who embeds trackers that use these techniques, we analyzed the sites' rank and category in our dataset which do so.
The ANOVA test shows that the category of a website strongly correlates with the usage of CNAME trackers ($p$-value $<0.0001$). 
These results are similar to findings of previous works (\eg\cite{Englehardt.OpenWPM.2016,Urban.WWW.2020}) that ``News'' websites tend to use more trackers -- this also holds for CNAME tracking.
In contrast, the visited site's rank does not show such a strong impact ($p$-value $\approx 0.006$).  Hence, popular and less popular sites embed third parties that use CNAME cloaking alike.
An analysis of the companies providing these tracking techniques is given in Section~\ref{sec:ecosystem}, which discusses the first-party cookie tracking ecosystem.

\subsubsection{Cloaked First-Party Cookies}
In this work, we focus on the usage of first-party cookies to track users.
Aside from the cloaking the tracking URL, one way to abuse CNAME cloaking for tracking is to place tracking cookies in the first-party context.
Therefore, we are interested in analysing if cloaked requests use the first-party context to store (tracking) cookies. 
Overall, we found \empirical{69,961} (\empirical{6.7}\%) of all first-party cookies that are set by such requests.
According to our definition of tracking cookies (see Section~\ref{sec:measure_1st_cookie}, \empirical{28\% (19,676)} of those cookies can be used to track users.
In contrast, \empirical{25,918} are session cookies and accordingly \empirical{43,367} are persistently stored but probably not used for tracking purposes.

Overall, the identified tracking cookies are utilized by \empirical{8,508 (74\%)} sites in our dataset.
Hence, it seems that the third parties actively try to circumvent anti-tracking methods. 
In our dataset, more popular sites (lower rank) and sites of a specific category (\eg ``News'') utilized CNAME based cookie tracking.
We provide a short overview of these distributions in Appendix~\ref{app:cname_cloaking}.
We provide a more detailed analysis of the sites and companies active in the first-party tracking ecosystem in Section~\ref{sec:ecosystem}.

For technical reasons, trackers that utilize CNAME cloaking are often included on a dedicated subdomain for which the DNS requests are redirected (\eg \texttt{tracking.foo.com}).
Hence, by default, the cookies would be set only for this origin.
However, we found that in \empirical{149,666 (95\%)} cases, the tracking cookies were set for the entire domain, which allows broader access to the cookie.

\subsubsection{Defense Mechanisms}
\emph{uBlock} provides tracking protection for CNAME cloaking based tracking as it actively inspects the DNS resolution of requests and makes decisions based on the result of it.
Other tools operate on the objects (\eg URLs) accessible on browser level and, therefore, cannot detect DNS level tracking approaches.
Therefore, we are interested in the effect of the `lower level' DNS blocking and the impact on users' privacy.
In our measurement, we used the \emph{uBlock} browser extension in profile \#3 and \#4.
In both profiles, we see a decrease of CNAME cloaking of around \empirical{46\%} what results in a decrease of \empirical{15\%} in identified trackers, which use this technology.
However, we still found trackers that were not detected by \emph{uBlock}, which is in line with previous work that showed that list based blockers miss some trackers~\cite{Merzdovnik.2017}.
The blocking of the identified requests resulted in a decrease of \empirical{101,387 (21\%)} first-party tracking cookies.
Our results show that modern anti-tracking tools need to adapt to modern first-party tracking techniques -- especially CNAME tracking.
On a technical side, current tools can add support to identify CNAME cloaking-based tracking if browser vendors provide an interface to the DNS resolution of the browser, which currently only Firefox does.

\subsubsection{Summary}
Our analysis  highlighted the scale of current first-party tracking in the wild.
While current technical solutions to protect users' privacy do not work for the novel tracking techniques, small adjustments in modern browser interfaces would allow the tools to do so.
In the previous sections, we have shown that first-party tracking is present on websites of different content and popularity.
However, the question arises who is providing the new tracking techniques and how the companies are connected in the ecosystem.

\subsection{First-Party Cookie Tracking Ecosystem}
\label{sec:ecosystem}
In the previous section, we have analyzed the extent of first-party tracking -- emphasizing first-party tracking cookies -- which are set through various means (\eg CNAME cloaking).
In the following, we shed some light on the ecosystem behind this first-party tracking and the companies that are active in it.
Concerning the websites that utilize first-party cookies for tracking, we found that \empirical{10.719 (91\%)} sites use them.
To get a better understanding of who uses such techniques, we use, on the one hand, the ranking (according to the \emph{Tranco} list~\cite{LePochat.Tranco.2019}) and, on the other hand, the category of the website's content (see Section~\ref{sec:measurement_framework}) to test for usage differences.
Again, we used the $X^2$ test ($\alpha = 0.5$) to evaluate  statistical significance in these categories.
Both categories (rank and category) show a strong correlation ($p$-value $< \empirical{0.0001}$) in the usage of first-party tracking cookies.
More specifically, we found that popular websites (rank $\le3,000$) and websites of categories like ``Business'', ''News'', or ``Education'' tend to use more of such tracking techniques.
The distribution along the ranks and categories can be found in Appendix~\ref{app:1st_party_cookies}.
This finding hints that websites are more aware of upcoming changes respectively current protection mechanisms and try to circumvent them.

\subsubsection{Companies Active in the Ecosystem}
As previously mentioned (see Section~\ref{sec:measurement_framework}), we use the \emph{Cookiepedia} to map identified cookies to companies using them.
Using this approach, we could classify \empirical{95.236 (36\%)} of the observed cookies.
These can be attributed to \empirical{283} distinct companies.
As previously briefly noted, the aggregated view of active companies, that use first-party tracking cookies, shows that \emph{Google} (\empirical{54\%}) and \emph{Facebook} (\empirical{14\%}) are the most common ones, followed by \emph{Adobe} (\empirical{8\%}) and \emph{OneTrust} (\empirical{3\%}). 
However, we see a different picture when we look at the companies to which the cookies are leaked.
\emph{Dynamic Yield} (17\%) and \emph{Adobe} (15\%) are the top companies that receive them, these results are in line with previous work~\cite{Dao.CNAME.2020}. 
\emph{Google} only receives 7\% of all leaked cookies.

As we have shown tracking via CNAME cloaking appears at large scale in the wild.
We use the \emph{WhoTracksMe} database~\cite{WhoTracksMe2018} to map the companies to the observed (CNAME cloaked) tracking domains.
Using this approach, we could classify \empirical{50\%} of the identified URLs.
These trackers can be attributed to \empirical{886} different companies.
Similar to the providers of first-party cookie trackers, \emph{Google} dominates the market of CNAME cloaking based tracking market (\empirical{39\%}). 
Other top players have a significant smaller market share (\emph{Facebook}: \empirical{6\%}, \emph{AdRoll}: \empirical{4\%}, and \emph{Amazon}: \empirical{3\%}).

\begin{figure*}[tb]
    \centering
    \includegraphics[width=\textwidth]{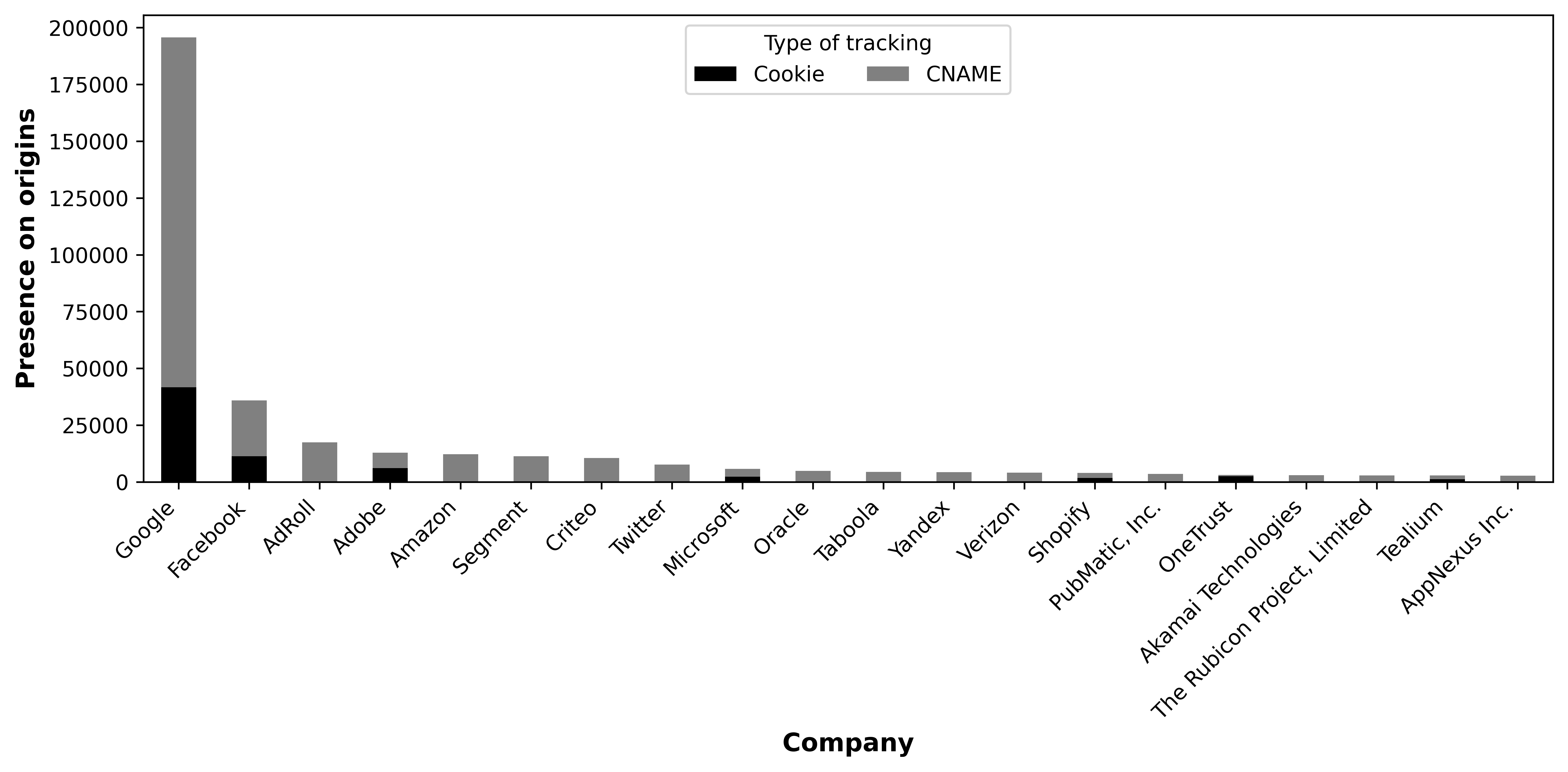}
    \caption{Companies active in the first-party tracking ecosystem.}
    \label{fig:ecosystem}
\end{figure*}

Figure~\ref{fig:ecosystem} provides a general overview of the companies active in the first-party tracking ecosystem.
In our analysis, in contrast to previous studies~\cite{Dao.CNAME.2020}, \emph{Google} and \emph{Facebook} also participate prominently in the CNAME tracking ecosystem.
Some companies exclusively relay on CNAME tracking as first-party tracking tool (\eg \emph{Amazon} or \emph{Criteo}) while others use a mix of them (\eg \emph{Adobe}). 

\subsubsection{Result Summary}
In this chapter, we show that the first-party tracking ecosystem is composed of the same players that are also active in the third-party tracking ecosystem (\eg \emph{Google}, \emph{Facebook}, or \emph{Amazon}).
Furthermore, we show that players who already dominate the third-party tracking market also do so in the first-party equivalent.
It is interesting to see that while on the one hand \emph{Google} aims to protect users' privacy by constraining the usage of third-party cookies~\cite{Google.Cookies.2020} at the same time uses techniques that stores and leaks identifiers in the first-party context.
A technique that is arguably harder to protect against.
Regarding the smaller players, we can see that companies often only use one of the analyzed techniques.

\section{Related Work}
\label{sec:related_work}
Online tracking has received a lot of attention from the academic community and the industry alike and, therefore, a huge body of work exists that analyzes tracking methods along different dimensions.
Quantifying Web tracking though measurement is a popular way to analyze different tracking techniques like device fingerprinting (\eg\cite{Englehardt.OpenWPM.2016,Vastel.18}), tracking cookies (\eg\cite{Acar.Tracking.2014,Englehardt.2015}), connections between tracking companies (\eg\cite{Papadopoulos.Cookie.2018}), other ways to track users (\eg\cite{Fouad.Missed.20}), and the efficiency of privacy-preserving tools to prevent racking (\eg\cite{Merzdovnik.2017,Roesner.2012}).
The impact of newly introduced  legislation on online tracking (like the CCPA or GDPR) was also analyzed in detail by different works (\eg\cite{Urban.Cookie.2020,sorensen2019before,Sanchez.OptOut.2019}).
However, all of these works focus on tracking performed by third parties that are embedded into a website.
Our work analyzes a new trend in the tracking market that moves the tracking code to the first party.

Recently, different works focused on CNAME cloaking.
In mid 2020, Dao \etAl were the first ones to analyze first-party tracking by performing a large scale Web measurement to identify the usage of CNAME cloaking~\cite{Dao.CNAME.2020}.
In their paper, they show that this technique has already been utilized for several years and they highlight the limitation of current privacy preserving technologies to counter such tracking.
Dao \etAl also provide a first mitigation strategy for CNAME based cloaking, using machine learning~\cite{dao2020machine}.
Complementary to our work, Ren \etAl take a look at the effect of CNAME cloaking on browser cookie policies and how they propagate in the ecosystem~\cite{Ren.CNAME.21}. 
They find that CNAME cloaking is often use to circumvent browser policies and that sensitive data is leaked by this bypass.
Dimova \etAl perform a rigorous analysis of CANME cloaking-based tracking and discuss privacy and security issues that arise by the usage of this technique~\cite{Dimova.CNAME.21}.
Finally, the dangers of CNAME cloaking is also discussed in several non-academic articles(\eg\cite{CNAME.Medium.2019,CNAME.Hackernews.2021}) and on browser vendor pages(\eg\cite{CNAME.Webkit.2020}), which hints the piratical relevance of the problem.
Most recently, Quan \etAl also took a look at cookie-based first-party tracking~\cite{Quan.Cookies.21}.
They use an elaborated JavaScript taint analysis framework to understand the process how first-party cookies are set and by whom.
They show that nearly all of the popular websites (97\%) utilize first-party cookies for user tracking.

Our approach differs from the named work because, on the one hand, we do not limit our analysis to one technique (\eg CNAME cloaking) only but analyse first-party tracking along different axis and take a broader look at the implications of first-party cookie tracking, sometimes enabled by CNAME cloaking (\eg by analysing defence mechanisms, the ecosystem, or temporal development).
Hence, we do not analyze novel tracking techniques in isolation but aim to understand how they are used in combination.
\section{Discussion and Conclusion}
\label{sec:conclusion}
\label{sec:discussion}
This paper analyzed the scale of (potential) first-party tracking techniques on the top 15,000 websites.
In contrast to previous work, we have shown that tracking is not exclusively implemented through third-party means but that first-party content is used for such purposes as well.
Hence, limiting or blocking third-party content on websites will not prevent that users will be tracked by third parties.
Our work shows that first-party cookies and requests that are redirected on DNS level are used for these purposes.
One takeaway from these findings is that privacy-preserving technologies need to extend their attempts to the first-party context while still covering third-party contexts.

First-party cookies that hold a personal identifier constitute a privacy risk since users might assume that those are exclusively used by the first party and are needed to run the service.
However, we have shown that they are often leaked to a third party, which could abuse them for tracking purposes.
While the extend of tracking and used methods (\eg first-party cookie syncing) are not yet present at scale we argue that they will gain in importance once first-party tracking is more commonly used.

Concerning CNAME based first-party tracking tracking, it is essential to highlight that uncovering such tracking attempts is almost impossible for everyday users or even experts because it is not observable on the application level.
Furthermore, service providers are well aware of this questionable interference in the users' privacy since they have actively to configure the redirection on their DNS server.
This practice is a substantial contrast to other tracking attempts where services ``rent out'' space on their services to third parties without knowing which content they will provide (\eg ads).

\subsubsection*{Acknowledgments}
This work was partially supported by the German Federal Ministry for Economic Affairs and Energy (grant 01MK20008E ``Service-Meister'' and grant 01MN21002H ``IDunion''), the Ministry of Culture and Science of North Rhine-Westphalia (MKW grant 005-1703-0021 ``MEwM'').
Furthermore, we would like to thank Norman Schmidt and Katharina Meyer for their efforts in developing the analysis pipeline.

\small
\bibliographystyle{plainurl}
\bibliography{bibliography}

\appendix
\section{Websites Utilizing  CNAME-based Cookie Tracking}
\label{app:cname_cloaking}
\begin{figure*}
    \centering
    \begin{subfigure}{0.45\columnwidth}
        \includegraphics[width=\linewidth]{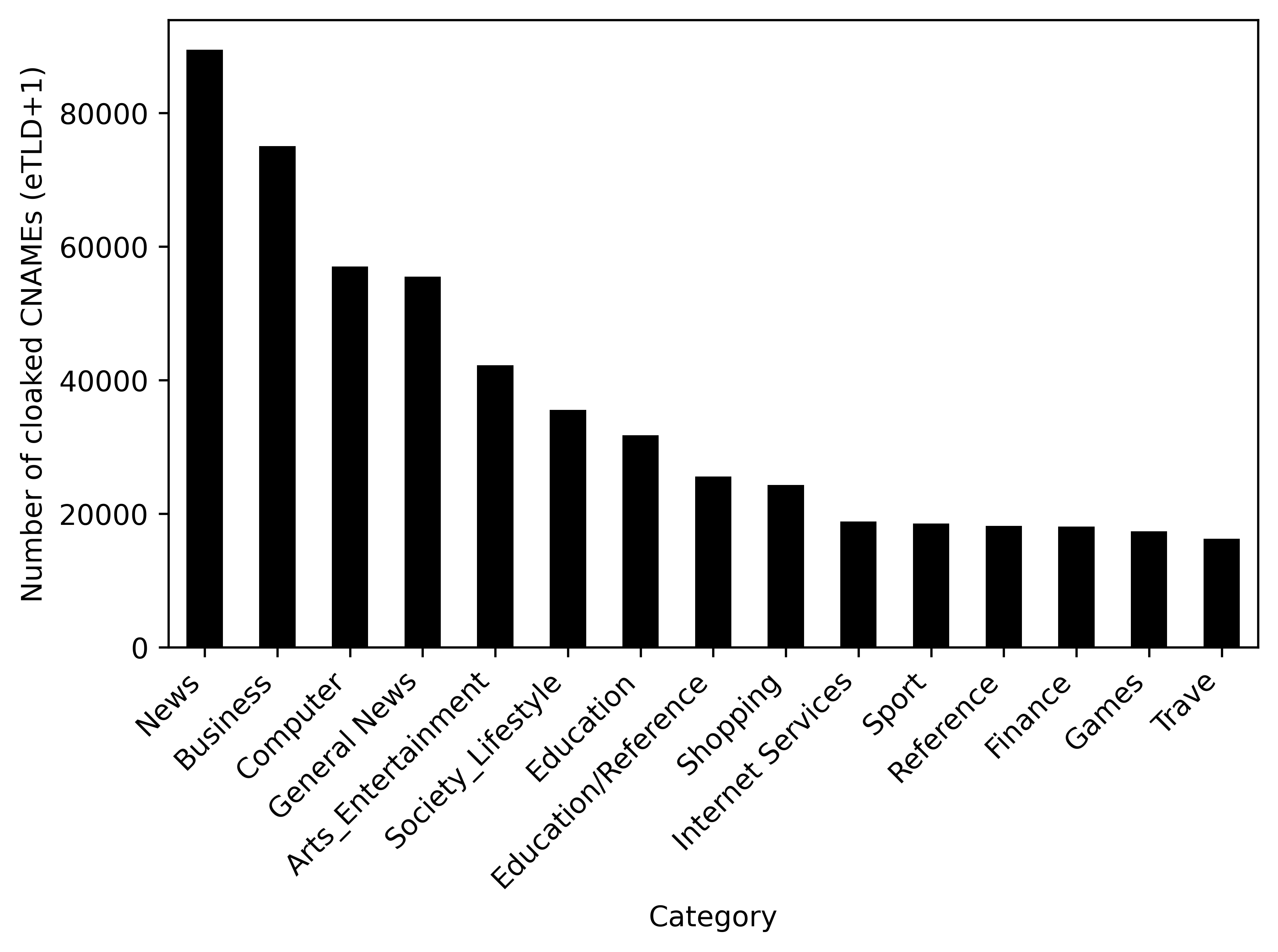}
        \caption{Category of websites that utilize CNAME cloaking.}
        \label{fig:cname_cat}
    \end{subfigure}
    \begin{subfigure}{0.45\columnwidth}
        \includegraphics[width=\linewidth]{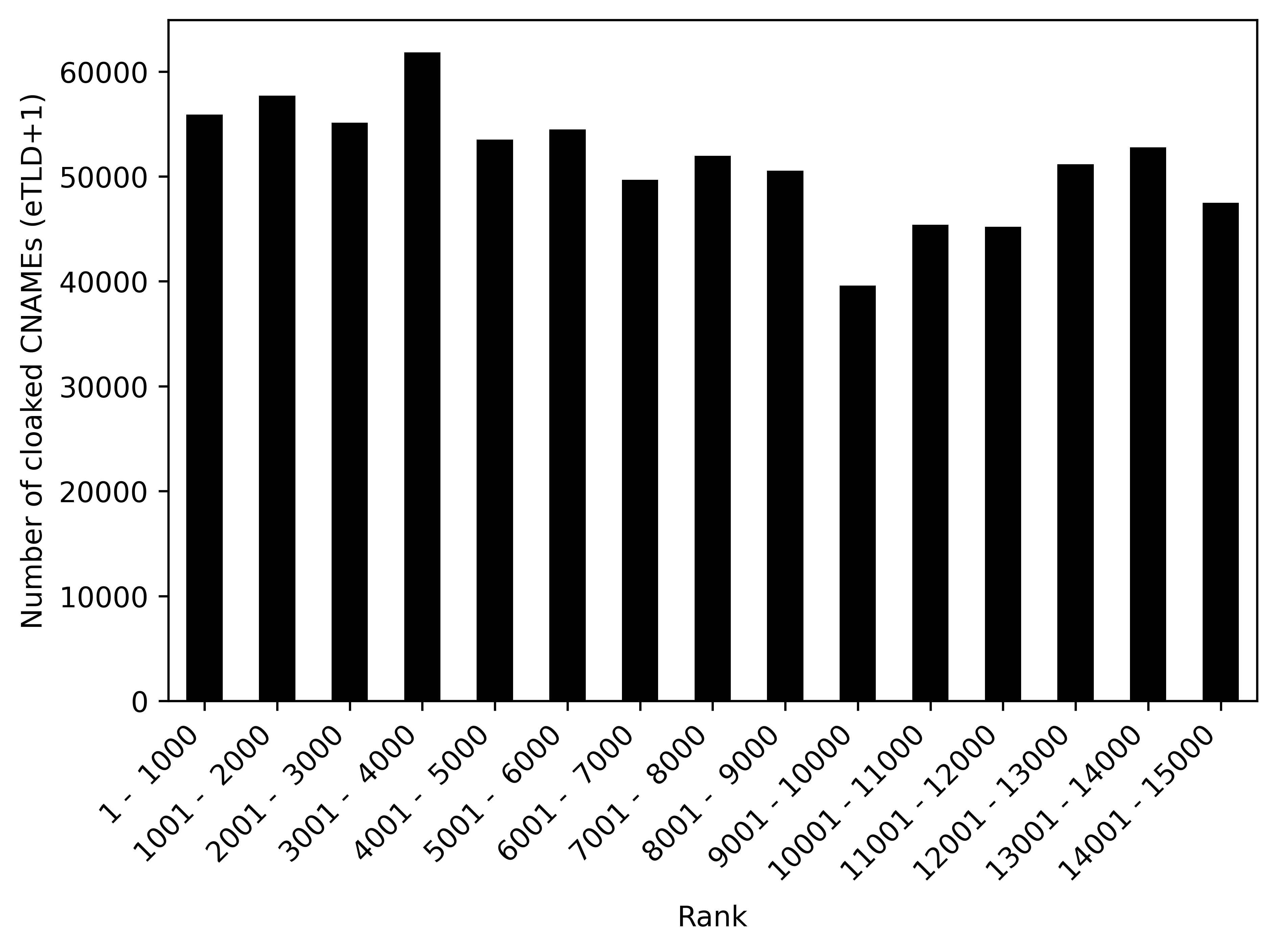}
        \caption{Rank of websites that utilize CNAME cloaking.}
        \label{fig:cname_rank}
    \end{subfigure}
    \caption{Comparison of the impact of the rank and category of a website to the usage of CNAME-based cookie tracking in terms of eTLD+1s that are cloaked.}
    \label{fig:CNAME_categories}
\end{figure*}

Websites that provide different content  different types of content and of all popularity levels use CNAME-based cookie tracking.
Figure~\ref{fig:cname_cat} shows the 'category' of each website, based on the 
categorisation of \emph{McAfee SmartFilter Internet Database}~\cite{mcaffee.2019}.
One can see that \emph{News} and \emph{Business} sites often use this type of tracking.
Furthermore, a non-negligible number of \emph{Education} websites tend to use it.
Figure~\ref{fig:cname_rank} highlights that the rank of a website has only little effect on the usage of the technique.

\section{Usage of First-Party Tracking Cookies}
\label{app:1st_party_cookies}
\begin{figure*}
    \centering
    \begin{subfigure}{0.45\columnwidth}
    \includegraphics[width=\linewidth]{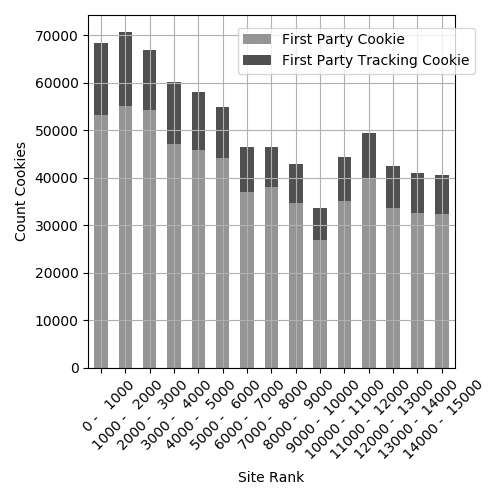}
        \caption{Profile \#1 and \#2}
        \label{fig:app:1st-party_cookie_distibution_1}
    \end{subfigure}
    \begin{subfigure}{0.45\columnwidth}
    \includegraphics[width=\linewidth]{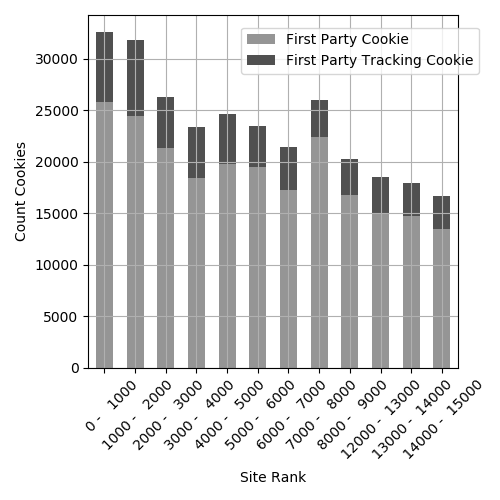}
        \caption{Profile \#3 and \#4 (\emph{uBlock active})}
        \label{fig:app:1st-party_cookie_distibution_2}
    \end{subfigure}
    \caption{Distribution of first-party cookies used to track users on the measured websites, by rank}
    \label{fig:app:1st-party_cookie_distibution}
\end{figure*}

First-party tracking via cookies in behalf of a third party is a common problem on the Web.
Figure~\ref{fig:app:1st-party_cookie_distibution_1} shows the categories of websites that uses such technique and Figure~\ref{fig:app:1st-party_cookie_distibution_2} shows this by rank.

\section{First-Party Tracking Over Time for all Websites}
\label{app:over_time}

\begin{figure}[H]
    \centering
    \includegraphics[width=0.8\columnwidth]{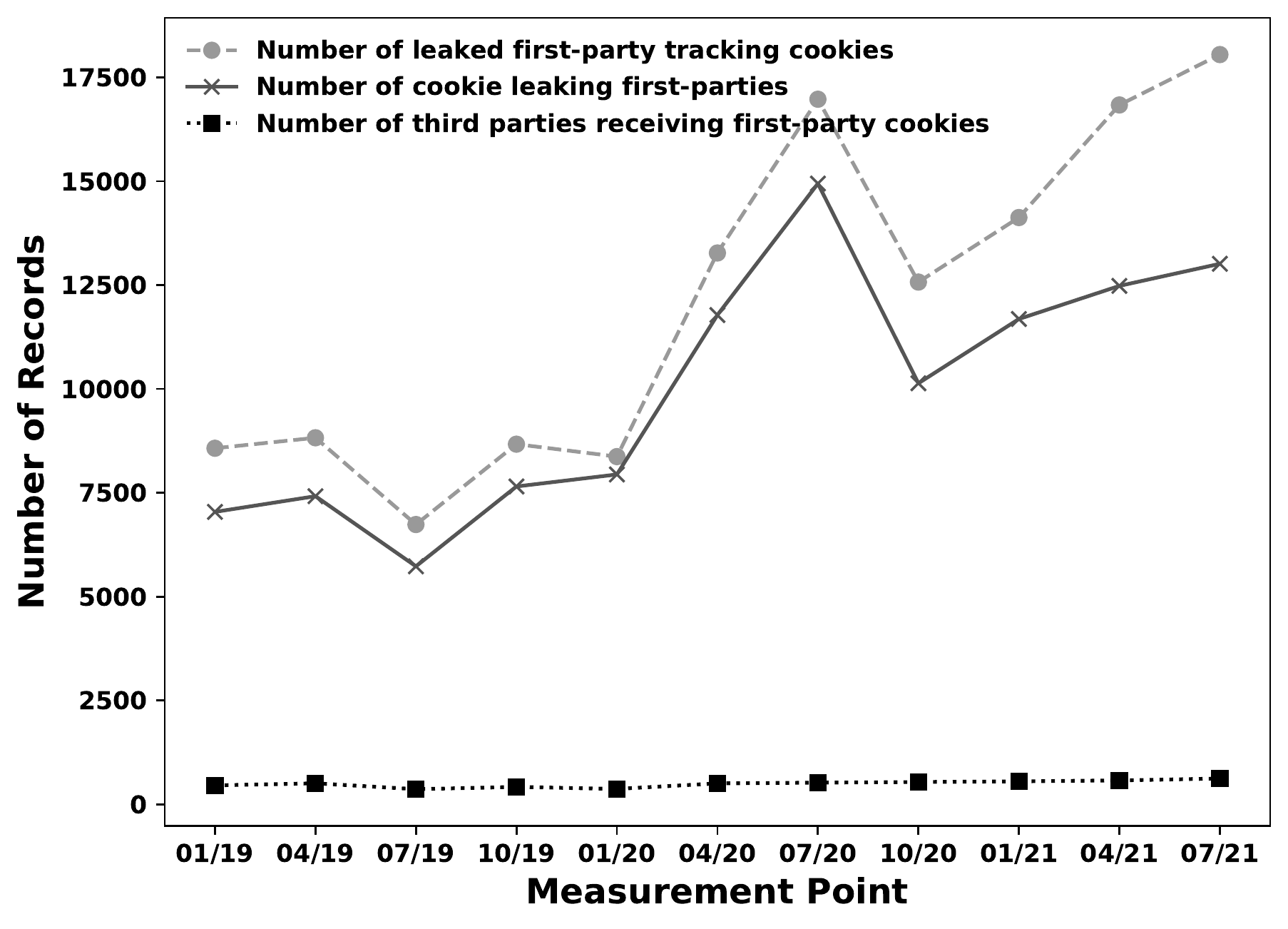}
    \caption{Number of distinct sites and third parties that engage in  first-party tracking and number of leaked first-party tracking cookies, over time -- for our entire dataset.}
    \label{fig:first_party_tracking_over_time_all_sites}
\end{figure}

Figure~\ref{fig:first_party_tracking_over_time_all_sites} provides an overview of the temporal development of first-party cookie based tracking on the Web.
In contrast to Figure~\ref{fig:first_party_tracking_over_time}, this figure shows the development for all websites in the dataset.
As one can see, the number of leaking parties and leaked cookies are directly related to each other.

\end{document}